\documentclass[11pt,a4paper]{article}
\pdfoutput=1
\usepackage{jheppub}
\usepackage{tikz}

\newcommand{\Tr}{\text{Tr}}

\usepackage{braket}

\title{Linearized Einstein's Equation around pure BTZ \\ from  Entanglement Thermodynamics}

\author[a,b]{Partha Paul,}
\author[c]{Pratik Roy.}

\affiliation[a]{Institute of Physics,\\ Sachivalaya Marg, Bhubaneshwar, India-751005}
\affiliation[b]{Homi Bhabha National Institute,\\ Training School Complex, Anushakti Nagar, Mumbai, India-400085}
\affiliation[c]{Chennai Mathematical Institute, \\ H1, SIPCOT IT Park, Siruseri 603103, India}

\emailAdd{pl.partha13@gmail.com}
\emailAdd{roy.pratik92@gmail.com}

\abstract{
It is known that the linearized Einstein's equation around the pure AdS can be obtained from the constraint $ \Delta S = \Delta\left< H \right> $, known as the first law of entanglement, on the boundary CFT. The corresponding dual state in the boundary CFT is the vacuum state around which the linear perturbation is taken. In this paper we revisit this question, in the context of $ \textnormal{AdS}_3/\textnormal{CFT}_2 $, with the state of the boundary $\textnormal{CFT}_2$ as a thermal state. The corresponding dual geometry is a planar BTZ black hole. By considering the linearized perturbation around this black brane we show that Einstein's equation follows from the first law of entanglement. } 

\begin{document}
\maketitle

\section{Introduction}

After the realization of a connection between gravity and thermodynamics \cite{bekenstein, bardeen, hawking} various attempts have been made to understand gravitational dynamics from horizon thermodynamics \cite{jacobson, padmanabhan, verlinde}. The discovery of AdS/CFT correspondence \cite{maldacena, ooguri} led to the new idea that the dynamics of spacetime can be understood from some sort of entanglement between the degrees of freedom of the boundary CFT \cite{mav, faulkner, myers, speranza, mosk, czech, swingle, ted, xian}. See also \cite{raamsdonk, raamsdonk1, raamsdonk2}. In this note, following \cite{mav, faulkner, myers}, we explore this idea further. 

Linear perturbations around a fixed reference state in the continuum field theory satisfy the first law of entanglement, $ \Delta S = \Delta \left< H \right> $ \cite{mav, faulkner, myers}, where $ S $ is the entanglement entropy of a spatial region and $ H  $ is the modular Hamiltonian associated with that region. In the AdS/CFT framework, each side of this first law can be computed using the dual geometry. In \cite{mav, myers}, the vacuum of a holographic CFT, with corresponding dual geometry pure AdS, was chosen as a fixed reference state. Considering perturbations around pure AdS, \cite{mav, myers} calculated $ \Delta S $ and $ \Delta\left<H\right> $ hologrpahically and showed that to linear order in the perturbation the first law of entanglement is satisfied, while inclusion of higher order contributions gives the constraint $ \Delta \left< H \right> \geq \Delta S $ \cite{mav, myers}\footnote{For related discussions one can also see, e.g, \cite{jb, numasawa, nozaki, allahbakhshi, wong} }. In \cite{mav, faulkner} linear perturbations were considered and it was shown that Einstein's equations linearized around pure AdS do follow from the first law of entanglement, thus showing their equivalence at first order.

In this paper, we take the thermal state of a holographic CFT as the fixed reference state and perturb it infinitesimally. The change in the entanglement entropy and the modular Hamiltonian of a spatial region will satisfy the first law of entanglement. Based on the holographic dictionary we then compute each side of this relation using metric components of the dual geometry. We show that for metric components of the dual geometry satisfying the linearized Einstein's equations, the first law of entanglement holds. Then we go the other direction, i.e, we show that the first law of entanglement fixes the metric uniquely if we demand that it holds in all frames of reference. 

 Entanglement entropy for a holographic field theory can be computed by applying Ryu-Takayanagi formula \cite{rt} and its covariant generalization \cite{hrt}. See also \cite{colgain} for some useful discussions on holographic enetanglement entropy in case of warped $ \textnormal{AdS}_3 $ geometries. Computing modular Hamiltonian in the field theory side is not an easy task. There are only few cases where it can be expressed as the integral of some local quantity, mainly the stress tensor \cite{casini,cardy}. The modular Hamiltonian for a spatial interval of a two dimensional CFT at finite temperature was calculated in \cite{casini}. Using the holographic dictionary one can obtain the boundary stress tensor from the asymptotic behavior of the metric components \cite{myersbst, skenderis, solodukhin}, which can then be used to compute modular Hamiltonian. See also \cite{erdmenger, bala, kostas1} for relevant discussions on holographic stress tensor.

\section{First Law of Entanglement $ \Delta S = \Delta \left< H \right> $}

In this section we briefly review the first law of entanglement. Relative entropy quantifies distinguishability between two states of a quantum field theory in the same Hilbert space \cite{ved}\footnote{ For relative entropy of excited states in two dimensional CFT see \cite{ugajin,jabbari}.}. Let us consider a $d$-dimensional Minkowski spacetime $ \mathbb{R}^{1,d-1} $. Also consider a spatial region $ A $ on a fixed time slice $ t=t_0 $ on this spacetime.
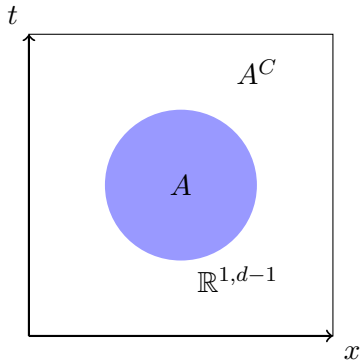
\begin{figure}
\center
\begin{tikzpicture}
\draw (0,0) rectangle (4,4);
\fill[blue!40!white] (2,2) circle (1cm) node[text=black](center) {$A$}; 
\node at (3,3.5) {$A^C$};
\node at (2.75,.75) {$\mathbb{R}^{1,d-1}$};
\draw[thick,->] (0,0) -- (4,0) node[anchor=north west] {$x$};
\draw[thick,->] (0,0) -- (0,4) node[anchor=south east] {$t$};
\end{tikzpicture}
\caption{Spatial region $A$ defined on a constant time slice $ t=t_0 $ of $d-$dimensional Minkowski spacetime $ \mathbb{R}^{1,d-1} $. $ A^C, $ the compliment of $A$, denotes the rest of the spacetime. $ \sigma_A $ denotes the reduced density matrix defined for this spatial region $A$, where $ \sigma $ is a state (represented as a density matrix) of a QFT defined on $ \mathbb{R}^{1,d-1} $.}
\end{figure}
 Let density matrices $ \sigma $ and $ \rho $ define two states of a QFT defined on $ \mathbb{R}^{1,d-1} $. Then the relative entropy of $\rho$ with respect to $\sigma$ for the spatial region $ A $ is defined as 
\begin{equation}\label{re}
S_A(\rho_A|\sigma_A) = -\Tr(\rho_A\log\sigma_A)-S_A(\rho_A)=\Tr(\rho_A \log \rho_A) - \Tr(\rho_A \log \sigma_A),
\end{equation}
where $ \rho_A $ and $ \sigma_A $ are the reduced density matrices associated with the spatial region $ A $.

Choose $\sigma$ as the reference state. With the definitions of entanglement entropy of a region $ A $, $ S_A(\rho_A) = -\Tr(\rho_A \log \rho_A)$, and of the modular Hamiltonian associated with that region, $ H_A = - \log \sigma_A $, one can rewrite ($ \ref{re} $) as 
\begin{equation}\label{re1}
S_A(\rho_A|\sigma_A) = \Delta \left< H_A \right> - \Delta S_A,
\end{equation}
where 
\begin{equation}\label{cee}
\Delta S_A = S_A(\rho_A) - S_A(\sigma_A),
\end{equation}
and
\begin{equation}\label{cmh}
\Delta \left< H_A \right> = \Tr(\rho_A H_A) - \Tr(\sigma_A H_A).
\end{equation}
It has  the nice positivity property that  $ S_A(\rho_A|\sigma_A) \geq 0$ with the inequality saturated if and only if $ \rho_A = \sigma_A $. Using this property one can show easily that for $ \rho_A $ very close to $ \sigma_A $ the relative entropy vanishes at linear order of perturbation giving rise to the constraint \cite{mav, myers}
\begin{equation}\label{fl}
\Delta S = \Delta \left< H \right>   .
\end{equation}
This is known as the first law of entanglement. 

Below, we will be interested in the entanglement entropy of a single spatial interval $ A=(-R,+R) $ for a holographic state of a CFT in 1+1-dimensions. We choose the initial holographic state to be a thermal state with temperature $ T=\frac{1}{2\pi} $. In this case the modular Hamiltonian is given by \cite{cardy, Hartman:2015apr}\footnote{We thank David Blanco for poiting out reference \cite{Hartman:2015apr} to us.}
\begin{equation}\label{tmh}
H_A = \frac{4\pi}{\sinh R} \int_{-R}^{+R} dx \left[\sinh\left(\frac{R+x}{2}\right)\sinh\left(\frac{R-x}{2}\right)\right]  T_{00}(x,t)
\end{equation}
where $ T_{00}(x,t) $ is the time-time component of the field theory stress tensor.
\begin{figure}        
\center
\begin{tikzpicture}
\draw[very thick,->] (0,0) -- (8,0) node[anchor=north west] {$x$};
\draw[thick] (2.5,-.2)  -- (2.5,.2) node[anchor=south] {{\small $-R$}};
\draw[thick] (5.5,-.2)  -- (5.5,.2) node[anchor=south] {{\small $R$}};
\draw[very thick,->] (0,-1) -- (0,1.5) node[anchor=south] {$t$};
\node at (4,.5) {{\large $A$}};
\node at (1,.5) {{\large $A^C$}};
\node at (7,.5) {{\large $A^C$}};
\end{tikzpicture}
\caption{Spatial region $A\equiv(-R,R)$ on a fixed time slice $ t=t_0 $ of the boundary spacetime $ \mathbb{R}^{1,1} $. Our modular Hamiltonian ($\ref{tmh}$) is defined for this region in a thermal state of a CFT with temperature $ T=\frac{1}{2\pi} $}
  \label{cftregion}
\end{figure}
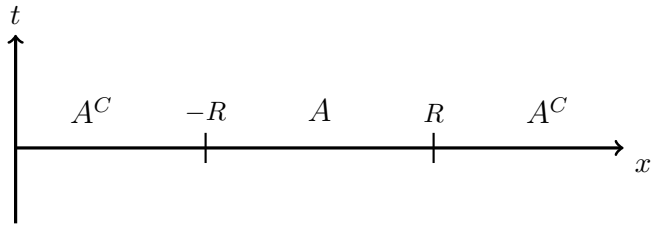


\section{Holographic Computation of $ \Delta S $ and $ \Delta \left< H \right> $}

We consider a thermal state of 1+1-dimensional holographic CFT with temperature $ T=\frac{1}{2\pi} $ as a fixed reference state. We know that the dual geometry of a thermal state of a holographic CFT is a black hole in AdS. In 2+1 bulk dimensions, it is the BTZ black hole. We would consider the BTZ black brane instead of black hole because the field theory is defined on $ \mathbb{R}^{1,1} $. The static BTZ black brane metric is\footnote{Our convention is that we use indices $ K,L $ for the three bulk coordinates $ \lbrace  z,t,x  \rbrace $ and indices $ \mu,\,\nu $ for two boundary coordinates $ \lbrace t, x \rbrace $.} \cite{btz}
\begin{equation}\label{btz}
ds^2 = \frac{1}{z^2}\left[ \frac{dz^2}{1-z^2} -(1-z^2)dt^2 + dx^2 \right]
\end{equation}
Here we have set the AdS radius of curvature, $L_{\textnormal{AdS}}=1$. With this convention one can check that the inverse temperature of this black brane is $ \beta=2\pi $.

\subsection{Holographic Computation of $ \Delta S $}

Let us consider a spatial interval $A=(-R, +R)$ at a fixed time $ t=t_0 $ in the boundary spacetime $ \mathbb{R}^{1,1} $. The entanglement entropy of this interval can be computed using the Ryu-Takayanagi formula \cite{rt}:
\begin{equation}\label{hee}
S_A = \frac{\textnormal{Length}(\gamma_A)}{4G_N},
\end{equation}
where $ \gamma_A $ is a geodesic in the bulk homologous to $ A $, and $ G_N $ is Newton's constant. When the spacetime is not static, one needs to use its covariant generalization \cite{hrt}. We will deal with that in the case of boosted black brane.

Let us now concentrate on our specific situation. The equation for the geodesic, with prescribed boundary conditions $ t=t_0=0,\ z=0, \ x=\pm R, $ is 
\begin{equation}\label{es}
z^2\cosh^2 R + \cosh^2 x = \cosh^2 R.
\end{equation}
Performing a coordinate transformation to Fefferman-Graham(FG) coordinates \cite{fg},
\begin{equation}\label{fg}
z^2 = \left( 1 + \frac{\tilde{z}^2}{4} \right)^{-2}\tilde{z}^2,
\end{equation}
we can write down the metric ($ \ref{btz} $) as
\begin{equation}\label{btzfg}
ds^2 = \frac{1}{\tilde{z}^2}\left[ d\tilde{z}^2 - \left( 1 - \frac{\tilde{z}^2}{4} \right)^2 dt^2 + \left( 1 + \frac{\tilde{z}^2}{4} \right)^2 dx^2  \right].
\end{equation}
 The advantage of writing down any asymptotically AdS spacetime in this coordinate is to identify the boundary stress tensor easily using the holographic prescription \cite{myersbst, solodukhin}. One can easily check that in case of pure BTZ ($ \ref{btzfg} $), the boundary stress tensor has the non-vanishing components $ T_{tt}=T_{xx}=\frac{1}{16\pi G_N} $. 
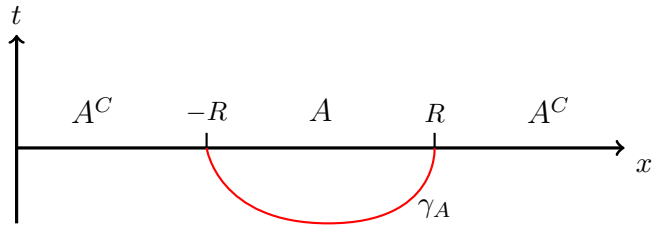
\begin{figure}       
\center
\begin{tikzpicture}                 
\draw[very thick,->] (0,0) -- (8,0) node[anchor=north west] {$x$};
\draw[thick] (2.5,0)  -- (2.5,.2) node[anchor=south] {{\small $-R$}};
\draw[thick] (5.5,0)  -- (5.5,.2) node[anchor=south] {{\small $R$}};
\draw[very thick,->] (0,-1) -- (0,1.5) node[anchor=south] {$t$};
\node at (4,.5) {{\large $A$}};
\node at (1,.5) {{\large $A^C$}};
\node at (7,.5) {{\large $A^C$}};
\draw [thick,red] plot [smooth,tension=1.75] coordinates{(2.5,0) (4.1,-1) (5.5,0)};
\node at (5.5,-.81) {$\gamma_A$};
\end{tikzpicture}
\caption{Pictorial representation of the Ryu-Takayanagi proposal. Area of the minimal area curve, i.e., length of the geodesic, $ \gamma_A $ gives the holographic entanglement entropy for the spatial region $A$ in the boundary.}
   \label{rtsurface}
\end{figure}
In terms of FG coordinates, the extremal surface equation ($ \ref{es} $) transforms as
\begin{align}
\frac{\tilde{z}^2}{\left( 1 + \frac{\tilde{z}^2}{4} \right)^2} =& \ \frac{\cosh^2 R - \cosh^2 x}{\cosh^2 R}        \nonumber    \\
\Rightarrow \qquad \qquad \tilde{z}^2 =& \ 4\frac{\cosh R - \cosh x}{\cosh R + \cosh x}  \ .    \label{esfg}
\end{align}
We can treat $ x $ as the intrinsic coordinate on the geodesic. The induced metric on the geodesic before perturbation is
\begin{equation}\label{ind}
g^{(0)}_{xx} = G^{(0)}_{KL}\frac{dx^K}{dx}\frac{dx^L}{dx},
\end{equation}
where $ G^{(0)}_{KL} $ are the metric components of pure BTZ ($ \ref{btzfg} $). The length functional is
\begin{equation}\label{area}
A = \int_{-R}^{+R} dx \ \sqrt{g^{(0)}_{xx}} .
\end{equation}


Now we add some pure metric perturbation to the BTZ metric ($ \ref{btzfg} $)\footnote{Linear perturbations around BTZ black brane were also considered in \cite{jyotirmay} and the first order correction to holographic entanglement entropy was calculated. They have also discussed the dynamics of the shift of holographic entanglement entropy.}. Any such perturbation in the FG coordinates can be written as
\begin{equation}\label{pert}
ds^2 = \frac{1}{\tilde{z}^2}\left[ d\tilde{z}^2 - \left( 1 - \frac{\tilde{z}^2}{4} \right)^2 dt^2 + \left( 1 + \frac{\tilde{z}^2}{4} \right)^2 dx^2+ \tilde{z}^2H_{\mu\nu}(\tilde{z},x,t)dx^{\mu}dx^{\nu} \right].
\end{equation}
From now on (in the case of static BTZ black brane) everything will be done in FG coordinates so we will drop the $\, \tilde{} \,$ sign over $\tilde{z}$ and simply write $z$. 

To linear order in perturbation $ H_{\mu\nu}(z,x,t) $, the change in the length functional (\ref{area}) is 
\begin{align}
\nonumber \Delta A =& \ \int dx \frac{1}{2}\sqrt{g^{(0)}}g^{(0)xx} \delta g_{xx},\\
=& \  \frac{1}{\sinh 2R}\int dx [\sinh(R+x)\sinh(R-x)]H_{xx}(z(x),x,t)  .         \label{carea}
\end{align}
Hence to first order in the perturbation, the change in the entanglement entropy is 
\begin{equation}\label{chee}
\Delta \hat{S}_A = \int dx [\sinh(R+x)\sinh(R-x)]H_{xx}(z(x),x,t).
\end{equation}
Here $ \Delta \hat{S}_A = 4 G_N \sinh (2R) \Delta S_A $.


\subsection{Holographic Computation of $ \Delta \left< H \right> $}

Equation ($ \ref{tmh} $) gives the modular Hamiltonian for a spatial interval $ A=(-R,+R) $ of a thermal state with temperature $ T=\frac{1}{2\pi} $. If we perturb this state infinitesimally, the change in the modular Hamiltonian is
\begin{equation}\label{cmh}
\Delta\left< H_A \right> = \frac{4\pi}{\sinh R} \int_{-R}^{+R} dx \left[\sinh\left(\frac{R+x}{2}\right)\sinh\left(\frac{R-x}{2}\right)\right]\Delta \left< T_{00}(x,t) \right> ,
\end{equation}
where $ \Delta \left< T_{00}(x,t) \right> $ is the change in the expectation value of the time-time component of the field theory stress tensor due to infinitesimal perturbations.

We have already mentioned that the boundary stress tensor can be found from the asymptotic form of the asymptotically AdS bulk metric. From \cite{myersbst, skenderis, solodukhin} we know that for a $d+1$-dimensional asymptotically AdS bulk metric written in FG coordinates, 
\begin{equation}\label{hstbtz}
ds^2 = \frac{1}{z^2}\left[ dz^2 + \eta_{\mu\nu}dx^{\mu}dx^{\nu} + z^d g_{\mu\nu}(z,x)dx^\mu dx^\nu \right],
\end{equation}
the boundary stress tensor can be found from the following asymptotic relation, 
\begin{equation}\label{hst}
\left< T_{\mu\nu}(x) \right> = \frac{d}{16\pi G_N}g_{\mu\nu}(z=0,x).
\end{equation}
Using the formula in our case, we have 
\begin{equation}\label{bst}
\left< T_{\mu\nu}(x,t) \right> = \frac{1}{8\pi G_N}\left( \frac{1}{2} + H_{\mu\nu}(z=0,x,t)\right),
\end{equation}
where $ \frac{1}{16\pi G_N} $ is the background boundary stress tensor. Hence the change in the modular Hamiltonian is 
\begin{equation}\label{chmh}
\Delta\braket{ \hat{H}_A} = 4\cosh R \int_{-R}^{+R} dx \left[\sinh\left(\frac{R+x}{2}\right)\sinh\left(\frac{R-x}{2}\right)\right]H_{tt}(x,t).
\end{equation}
where $ \Delta \braket{ \hat{H}_A} = 4 G_N \sinh (2R) \Delta \left<H_A\right> $.


\section{Proof That Einstein's Equations Imply $ \Delta S = \Delta \left<H\right> $}

We now have the expressions for both $ \Delta S $ and $ \Delta \left< H \right>  $ to linear order in the bulk perturbation. In this section we will show that the solutions of linearized Einstein's equations satisfy the relation $ \Delta S = \Delta \left<H\right> $.  \\

In $d+1$-dimensions with a cosmological constant $ \Lambda=-\frac{d(d-1)}{2L_{\textnormal{AdS}}^2} $, Einstein's equations read (recall that we have set $L_{\textnormal{AdS}}=1$) 
\begin{equation}\label{dEE}
R_{AB} - \frac{1}{2}G_{AB}(R+d(d-1)) = 0.
\end{equation}
Using the metric ($ \ref{pert} $) to linear order in $ H_{\mu\nu}, $ different components of equations ($ \ref{dEE} $) (with $d=2$) read 
\begin{align}
32 z H_{tt}(z,x,t) - (4-z^2)\left[-(12+z^2) \partial_z H_{tt}(z,x,t) - z (4-z^2) \partial_z^2 H_{tt}(z,x,t)\right] =& \ 0     \label{EE1}     \\
-32 z H_{xx}(z,x,t) + (4+z^2)\left[(12-z^2) \partial_z H_{xx}(z,x,t) + z (4+z^2) \partial_z^2 H_{xx}(z,x,t)\right] =& \ 0     \label{EE2}     \\
(48 + z^4)\partial_z H_{tx}(z,x,t) + z(16 - z^4)\partial_z^2 H_{tx}(z,x,t) =& \ 0      \label{EE3}     
\end{align}
\begin{align}
 -2& (4-z^2)^2 \partial_t H_{tx}(z,x,t) + 2 (16+z^4) \partial_x H_{tt}  \nonumber \\
 & \ \qquad \qquad \qquad \ - z (16-z^4)\,\partial_z \left(\partial_t H_{tx}(z,x,t) - \partial_x H_{tt}(z,x,t)\right) = \ 0      \label{EE4}     \\
2& (4+z^2)^2 \partial_x H_{tx}(z,x,t) - 2 (16+z^4) \partial_t H_{xx}   \nonumber \\
 & \ \qquad \qquad \qquad \ - z (16-z^4)\,\partial_z \left(\partial_t H_{xx}(z,x,t) - \partial_x H_{tx}(z,x,t)\right) = \ 0        \label{EE5}     
\end{align}
\vspace{-1.1cm}
\begin{align}
\nonumber  -2(4+z^2)^2 H_{tt}(z,x,t) + 2 (4-z^2)^2 H_{xx}(z,x,t) + z \left[16 z \partial_t^2 H_{xx}(z,x,t)   \right.  & \ \\
\nonumber  - 32 z \partial_t \partial_x H_{tx}(z,x,t) + 16 z \partial_x^2 H_{tt}(z,x,t) - (16-z^4)\partial_z H_{tt}(z,x,t)& \ \\
\left. + (16-z^4)\partial_z H_{xx}(z,x,t)\right] =& \ 0        \label{EE6}
\end{align}
Setting $z=0$ in ($ \ref{EE6} $) one can see that 
\begin{equation}\label{trace}
-H_{tt}(z=0,x,t) + H_{xx}(z=0,x,t) = 0.
\end{equation}
This holds because the boundary theory is a conformal field theory and ($ \ref{trace} $) is the tracelessness condition of the boundary field theory stress tensor. Now demanding the smoothness condition at $z=0$ and using ($ \ref{trace} $) one arrives at the following solutions for the perturbations: 
\begin{align}
H_{tt}(z,x,t) =& \ (4-z^2)H(x,t),       \label{shtt}        \\
H_{xx}(z,x,t) =& \ (4+z^2)H(x,t),       \label{shxx}        \\
H_{tx}(z,x,t) =& \ h(t,x)               \label{shtx}
\end{align}
with $ h(x,t) $ and $ H(x,t) $ restricted by the conditions,
\begin{align}
\partial_t h(x,t) =& \ 4 \partial_x H(x,t),     \\
\partial_x h(x,t) =& \ 4 \partial_t H(x,t),     \\
(\partial_t^2 - \partial_x^2 )H(x,t) =& \ 0                            
\end{align}
Now we consider the expressions for $ \Delta \hat{S}_A $ and $ \Delta \braket{\hat{H}_A}$. 
\begin{align}\label{fcs}
\nonumber \Delta \hat{S}_A   =& \   \int_{-R}^R dx \sinh(R+x)\sinh(R-x) H_{xx}(z,x,t_0)\\
\nonumber   =& \   \int_{-R}^R dx \sinh(R+x)\sinh(R-x) (4+z^2) H(x,t_0)\\
  =& \   16 \cosh R \int_{-R}^{+R} dx \sinh\left(\frac{R+x}{2}\right)\sinh\left(\frac{R-x}{2}\right) H(x,t_0)
\end{align}
and  
\begin{align}\label{fch}
\nonumber \Delta \braket{\hat{H}_A}   =& \   4\cosh R \int_{-R}^R dx \sinh\left(\frac{R+x}{2}\right)\sinh\left(\frac{R-x}{2}\right) H_{tt}(0,x,t_0)\\
  =& \   16\cosh R \int_{-R}^R dx \sinh\left(\frac{R+x}{2}\right)\sinh\left(\frac{R-x}{2}\right)  H(x,t_0)
\end{align}
Comparing ($ \ref{fcs} $) and ($ \ref{fch} $) we see that indeed $ \Delta S_A = \Delta \left<H_A\right> $.


\section{Proof That First Law of Entanglement Implies Einstein's Equations}

In this section we go the other direction. We will show that the constraint ($ \ref{fl} $) with the boundary condition ($ \ref{bst} $) fixes the metric uniquely. We would follow the strategy of \cite{mav} for the proof .        \\

Let $ H^{EE}_{\mu\nu} $ be the metric that solves Einstein's equations ($\ref{EE1}-\ref{EE6}$) with the boundary conditions ($ \ref{bst} $). We will show that there is no metric other than $ H^{EE}_{\mu\nu} $ with the boundary conditions ($ \ref{bst} $) which satisfies the same relation ($ \ref{fl} $). Suppose we have another metric $ H_{\mu\nu} $ satisfying ($ \ref{fl} $) with the same boundary condition ($ \ref{bst} $). We will show that
\begin{equation}\label{delmn}
\Delta_{\mu\nu}(z,x,t) \equiv H^{EE}_{\mu\nu} - H_{\mu\nu} = 0,
\end{equation} 
for all $z$. Just by demanding that both the metrics satisfy the same boundary condition, we already have 
\begin{equation}\label{delmn0}
\Delta_{\mu\nu}(0,x,t) = 0.
\end{equation}
Hence ($ \ref{chmh} $) tells us that 
\begin{equation}\label{cmh0}
\Delta(\Delta\left< H_A \right>) = 0.
\end{equation}
Now the constraint ($ \ref{fl} $) together with the expression ($ \ref{chee} $) tells us that in a fixed frame of reference (say, the $ t=t_0 $ frame) we have 
\begin{equation}\label{delxx0}
0 = \int{dx\, \Delta_{xx}(z(x),x+x_0,t_0)[\sinh(R+x)\sinh(R-x)]}.
\end{equation}
In the integral ($ \ref{delxx0} $) we have shifted the origin from $ x=0 $ to $ x=x_0 $.
Now we expand $ \Delta_{\mu\nu} $ as 
\begin{align}\label{expan}
\nonumber \Delta_{\mu\nu}(z(x),x+x_0,t_0)   =& \   \sum_{n=0}^{\infty}z^n \Delta^{(n)}_{\mu\nu}(x+x_0,t_0)\\
  =& \   \sum_{n,m} z^n \frac{x^{2m}}{(2m)!}\partial_x^{2m}\Delta^{(n)}_{\mu\nu}(x_0,t_0) .
\end{align}
Equation ($ \ref{delxx0} $) thus becomes 
\begin{equation}
\sum_{n,m}\frac{1}{(2m)!}\partial_x^{2m} \Delta^{(n)}_{xx}(t_0,x_0)\int_{-R}^{+R}dx [z^n x^{2m} \sinh(R+x)\sinh(R-x)] = 0.
\end{equation}
Substitung for $ z $ the expression ($ \ref{esfg} $) and performing some simplifications, we finally get 
\begin{equation}\label{eqn}
\sum_{n,m}\frac{2^{n+1}}{(2m)!}\partial_x^{2m} \Delta^{(n)}_{xx}(t_0,x_0) I_{n,m}(R) = 0,
\end{equation}
where we write
\begin{equation}\label{int}
I_{n,m} = \int_{0}^{R}dx \left[x^{2m}(\cosh R + \cosh x)^{-\frac{n}{2}+1}(\cosh R - \cosh x)^{\frac{n}{2}+1}\right].
\end{equation}
We now need to expand ($ \ref{eqn} $) in powers of $ R $ and set each coefficient to zero to see what constraints do they impose on $ \Delta_{xx}(t_0,x_0) $. Thus expanding ($ \ref{eqn} $) around $ R=0 $, we get 
\begin{align}\label{exeqn}
\sum_{n,m,j}\frac{2^{n+1}}{(2m)!}\partial_x^{2m} \Delta^{(n)}_{xx}(t_0,x_0)\frac{R^{j}}{j!}\left[\partial_R^j I_{n,m}(R)\right]_{R=0} = 0 .
\end{align}
Vanishing of the $ R^{J+3} -$th term requires 
\begin{align}\label{Delta}
\sum_{n,m}\frac{2^{n+1}}{(2m)!}\partial_x^{2m} \Delta^{(n)}_{xx}(t_0,x_0)\left[\partial_R^{J+3} I_{n,m}(R)\right]_{R=0} = 0 
\end{align}
The LHS of ($ \ref{Delta} $) contains a summation over $ n,\,m $ for a fixed $ J $. In appendix \ref{app} we explicitly show that for both odd and even $ J $ (for odd $ J $ numerical analysis is presented in appendix \ref{app}), all the terms in the summation vanish except when $ n \leq J $. For $ n=J $, we have \ $ m=0 $ and for $ n < J $, we need \ $ m \neq 0 $ (Please see appendix \ref{app} for details). The last non-vanishing term in the summation ($ \ref{Delta} $) with a fixed $ J $ is $ n=J, \ m=0. $ This term can be expressed as a linear combination of the lower order terms which establishes the result that 
\begin{equation}\label{Delta=0}
\Delta^{(J)}_{xx}(t_0,x_0) = 0
\end{equation}
for $J=0,1,2,\cdots$


Having shown that the entanglement first law fixes the solution to linearized Einstein's equations in the rest frame, we now consider a boosted frame and try to repeat the analysis above. Consider then the boosted BTZ black brane. The coordinate transformations to go from the static BTZ black brane to the boosted BTZ black brane are 
\begin{align}\label{lt}
t = \gamma(t^{\prime}-\beta x^{\prime}),    \\
x = \gamma(x^{\prime}-\beta t^{\prime}).
\end{align}
Here $ \beta, \gamma $ have their usual meanings from the special theory of relativity. The metric of the boosted black brane is given by 
\begin{equation}\label{bbtz}
ds^2 =  \frac{1}{z^2}\left[ \frac{dz^2}{1-z^2} + (-1 + \gamma^2 z^2){dt^{\prime}}^2 + (1 + \beta^2 \gamma^2 z^2){dx^{\prime}}^2 - 2\beta\gamma^2 z^2 dt^{\prime}dx^{\prime} \right]
\end{equation}
Notice that here, $z$ denotes the original radial coordinate and not the FG coordinate $\tilde{z}$. Below, we will explicitly write $ \tilde{z} $ for the FG coordinate. We need to solve the spacelike geodesic equations in this geometry with boundary conditions $ z=0,\,t=0,\,x=\pm R $. Working in the geometry ($ \ref{bbtz} $) with these boundary conditions is equivalent to working in the geometry ($ \ref{btz} $) with the following boundary conditions 
\begin{align}\label{bbc}
\nonumber z= 0, \quad  t=  -\beta \gamma R, \quad  x=  \gamma R, \\
z=  0, \quad t=  \beta \gamma R, \quad  x=  -\gamma R.
\end{align}
 Hence we will solve for the spacelike geodesics with metric ($ \ref{btz} $) and boundary conditions ($ \ref{bbc} $). Let $s$ be the proper length along the geodesic. Then we have the following equations for the spacelike geodesics 
\begin{align}
\frac{1}{z^2}\left[ \frac{1}{1-z^2}\left(\frac{dz}{ds}\right)^2 - (1 - z^2)\left(\frac{dt}{ds}\right)^2 + \left(\frac{dx}{ds}\right)^2 \right] =& \ 1,     \label{sg1} \\
\frac{1}{z^2}\left[ \frac{dx}{ds} \right] =& \ p,       \label{sg2}     \\
-\frac{1}{z^2}\left[ (1 - z^2)\left(\frac{dt}{ds}\right) \right] =& \ e  ,      \label{sg3}
\end{align}
where $ p $ and $ e $ are two constants of motion along the geodesic. We want to write down the geodesic in a parametric form $ z=z(x),~t=t(x) $. Hence we write down the above three equations in the following form 
\begin{align}
pz\frac{dz}{dx} =& \ \sqrt{1-(1-e^2+p^2)z^2 + p^2 z^4} \,  ,    \label{sgx1}        \\
\frac{dt}{dx} =& \ -\frac{e}{p(1 - z^2)} \,   .    \label{sgx2}
\end{align}
With the prescribed boundary conditions ($ \ref{bbc} $), the solutions are 
\begin{equation}\label{best}
\tanh(\gamma R)\tanh t + \tanh(\beta \gamma R)\tanh x = 0,
\end{equation}
and 
\begin{equation}\label{besz}
\cosh^2(\gamma R)z^2 + \left[ 1 - \frac{\sinh^2(\beta \gamma R)}{\sinh^2(\gamma R)} \right]\cosh^2 x = \left[ 1 - \frac{\sinh^2(\beta \gamma R)}{\sinh^2(\gamma R)} \right]\cosh^2(\gamma R).
\end{equation}
In terms of the FG coordinates,  ($ \ref{besz} $) becomes 
\begin{equation}\label{beszfg}
\tilde{z}^2 = 4\frac{\cosh(\gamma R) - \sqrt{(1-r)\cosh^2(\gamma R) + r\cosh^2 x}}{\cosh(\gamma R) + \sqrt{(1-r)\cosh^2(\gamma R) + r\cosh^2 x}},
\end{equation}
where 
\[
r = \left[ 1 - \frac{\sinh^2(\beta \gamma R)}{\sinh^2(\gamma R)} \right].
\]
The induced metric on the spacelike geodesic is 
\begin{equation}\label{bind}
ds^2_{ind} = \frac{\sinh^2(2\gamma R)}{[\cosh(2\gamma R)-\cosh(2x)]^2} dx^2.
\end{equation}
Change in the geodesic length is
\begin{align}\label{bcaf}
\nonumber \Delta A = \int_{-\gamma R}^{+\gamma R} dx\, \frac{\sinh(\gamma R + x)\sinh(\gamma R - x)}{\sinh(2\gamma R)} \left[ \left( \frac{dt}{dx} \right)^2 H_{tt} + 2\left( \frac{dt}{dx} \right)H_{tx}+ H_{xx}\right].
\end{align}
An argument similar to that leading to  ($ \ref{delxx0} $) gives the following equation for the boosted case 
\begin{equation}\label{b0}
\int_{-\gamma R}^{+\gamma R} dx \sinh(\gamma R + x)\sinh(\gamma R - x) \left[\left(\frac{dt}{dx}\right)^2 \Delta_{tt} + 2\left(\frac{dt}{dx}\right)\Delta_{tx} + \Delta_{xx} \right] = 0.
\end{equation}
Now assuming that $ \Delta_{\mu\nu}(z,x,t) $ is an analytic function, we can expand it in the following form 
\begin{align}\label{bexpan}
\nonumber \Delta_{\mu\nu}(\tilde{z},t+t_0,x+x_0)   =& \   \sum_{n=0}^{\infty}\tilde{z}^n\Delta_{\mu\nu}^{(n)}(t+t_0,x+x_0),\\
\nonumber   =& \   \sum_{n=0}^{\infty}\tilde{z}^n \left[\sum_{m_t,m_x}\frac{\partial_t^{m_t}\partial_x^{m_x}\Delta_{\mu\nu}^{(n)}(t_0,x_0)}{m_t! m_x!} \right]t^{m_t}x^{m_x}, \\
  =& \   \sum_{n,m_t,m_x} B^{n,m_t,m_x}_{\mu\nu} \tilde{z}^n t^{m_t} x^{m_x}.
\end{align}
 ($ \ref{b0} $) then becomes  
\begin{equation}\label{bsum}
\sum_{n,m_t,m_x} \left[ B^{n,m_t,m_x}_{tt} I^{n,m_t,m_x}_{tt}(R) + B^{n,m_t,m_x}_{tx} I^{n,m_t,m_x}_{tx}(R) + B^{n,m_t,m_x}_{xx} I^{n,m_t,m_x}_{xx}(R) \right] = 0,
\end{equation}
where 
\begin{align}
I^{n,m_t,m_x}_{tt}(\gamma R) =& \ \int_{-\gamma R}^{+\gamma R} dx  x^{m_x}[\sinh(\gamma R + x)\sinh(\gamma R - x)] \tilde{z}^n t^{m_t} \left(\frac{dt}{dx}\right)^2, \label{btt} \\
I^{n,m_t,m_x}_{tx}(\gamma R) =& \ 2\int_{-\gamma R}^{+\gamma R} dx   x^{m_x}[\sinh(\gamma R + x)\sinh(\gamma R - x)]\tilde{z}^n t^{m_t}  \left(\frac{dt}{dx}\right), \label{btx} \\
I^{n,m_t,m_x}_{xx}(\gamma R) =& \ \int_{-\gamma R}^{+\gamma R} dx x^{m_x}[\sinh(\gamma R + x)\sinh(\gamma R - x)]\tilde{z}^n t^{m_t}. \label{bxx}
\end{align}
We now use the same technique of expanding around $R=0$ as before, but the calculation is difficult even for even integer $n$. One can examine term by term the expansion series ($ \ref{bsum} $) and check which coefficients in a particular term are non zero. In the appendix \ref{app} we discuss the first few terms. Working in this manner we finally arrive, apart from some constant factors, at the following equations 
\begin{align}
\nonumber \Delta^{(0)}_{xx}-2\beta \Delta^{(0)}_{tx} + \beta^2 \Delta^{(0)}_{tt}   =& \   0\\
\nonumber \Delta^{(1)}_{xx}-2\beta \Delta^{(1)}_{tx} + \beta^2 \Delta^{(1)}_{tt}   =& \   0\\
 \Delta^{(2)}_{xx}-2\beta \Delta^{(2)}_{tx} + \beta^2 \Delta^{(2)}_{tt}   =& \   0     \label{beqns}       \\
 \vdots  \qquad \qquad & \     \nonumber
\end{align}
One can take any arbitrary value of $ n $ and check that the pattern should be the same, although for high values of $ n $ computations become difficult. Now ($ \ref{beqns} $) is a polynomial in $ \beta $ and the coefficients of each $\beta^{k}$ must vanish individually. From the coefficients of $ \beta^0, $  we get:
\begin{equation}\label{be0}
\Delta^{(n)}_{xx} = 0,
\end{equation}
from the coefficients of  $ \beta $:
\begin{equation}\label{be1}
\Delta^{(n)}_{tx} = 0,
\end{equation}
and from the coefficients of $ \beta^2 $:
\begin{equation}\label{be2}
\Delta^{(n)}_{tt} = 0,
\end{equation}
where $ n=0,1,2,\ldots $

Thus we have shown that $ (\ref{delmn}) $ is true for all $ z $. Note that we have assumed that $ \Delta_{\mu\nu}(z,x,t) $ is an analytic function. 

\section{Discussion}

In this paper we have shown that linearized Einstein's equations around the BTZ black brane can be obtained from the first law of entanglement thermodynamics, $ \Delta S = \Delta \left< H \right> $, where the reference state was taken to be a thermal state of the CFT which is dual to the black brane. It would be interesting to check if non-linear Einstein's equations can be obtained from some constraints on the entanglenemnt entropy of the thermal state of boundary CFT as well. In particular, in \cite{thomas} the vacuum state of the boundary CFT was perturbed by some scalar primary or stress tensor operators and it was shown that for such excited states, up to second order in the perturbation, the entanglement entropy  of all ball-shaped regions can be obtained using the covariant prescription for holographic entanglement entropy from the corresponding dual geometries. It was shown that the corresponding dual spacetimes must satisfy  Einstein's equations up to second order in the perturbation around AdS. It would be interesting to extend their work for the thermal state of holographic CFTs.  \\

One of the important points we would like to emphasize is that the Einstein equations that we have derived hold outside the horizon. This is because the spacelike geodesics  used to compute the entanglement entropy do not see the region behind the horizon. It will be very interesting if a similar method can be used to prove Einstein's equations behind the horizon. It seems that generalization  of this method to the two-sided eternal BTZ black hole may give us some insights into the derivation of Einstein's equations behind the horizon.\\

 It would also be interesting to check it for higher dimensional black holes and also at non-linear order. We have only considered the metric perturbations in the bulk such that only the boundary stress tensor has a non-trivial expectation value. One can also consider other excitations as well, e.g., turning on scalar field in the bulk where the scalar operator will acquire non-trivial expectation values in the boundary theory.
\\

{\noindent
\textbf{Note Added:} While this article was under preparation, the work \cite{Blanco:2018riw} appeared in the arXiv.\footnote{We are grateful to David Blanco for bringing this paper to our attention.} Some of their results, although they take a different approach, appear to agree with ours where they overlap.}

\section*{Acknowledgements}

We are grateful to Shamik Banerjee for suggesting this problem and guiding us throughout the project. We are also thankful to him for carefully reading the draft and for correcting some parts of it. We are thankful to Amitabh Virmani and Jyotirmay Bhattacharya for helpful discussions on related matters. We would also like to acknowledge the hospitality at Chennai Mathematical Institute during the workshop `Student Talks in Trending Topics in Theory (ST$^4$)'  where part of this work was done. PR is grateful to Debangshu Mukherjee for his generous help and multiple discussions on this work. PR is partially supported by a grant to CMI from the Infosys Foundation.


\appendix       

\section{ Analysis to show that $ \Delta^{(n)}_{\mu\nu}(t_0,x_0) = 0 $ for all $ n $}   \label{app}

The $j$-th derivative of $ I_{n,m}(R) $ ($ \ref{int} $) gives 
\begin{align}\label{jth}
 \partial_R^j I_{n,m}(R) =  \sum_{i=0}^{j-1}\partial_R^{j-i-1}\left( \partial_R^i f_{n,m}(x,R)|_{x=R} \right),
+\int_0^R \partial_R^j f_{n,m}(x,R) dx
\end{align}
where 
\begin{equation}
f_{n,m}(x,R) = x^{2m}(\cosh R + \cosh x)^{-\frac{n}{2}+1}(\cosh R - \cosh x)^{\frac{n}{2}+1}.
\end{equation}
Thus from ($ \ref{Delta} $) we have 
\begin{align}\label{sum}
\sum_{n,m}\frac{2^{n+1}}{(2m)!}\partial_x^{2m} \Delta^{(n)}_{xx}(t_0,x_0)\left[ \sum_{i=0}^{J+2}\partial_R^{J-i+2}\left( \partial_R^i f_{n,m}(x,R)|_{x=R}\right) \right]_{R=0} = 0.
\end{align}
First we will show that $\Delta_{xx}=0$ for even integers. For odd integers, we check it numerically below. Writing $2l$ for the even integer $n$, and  simplifying a bit, we get
\begin{equation}\label{jth0}
\sum_{l=0}^{J+1}\sum_{m}\frac{(l+1)!\,2^{l}}{(2m)!}\partial_x^{2m} \Delta^{(2l)}_{xx}(t_0,x_0)C^{J,l,m}(R=0) = 0,
\end{equation} 
where 
\begin{equation}\label{c}
C^{J,l,m}(R=0) = \left[\partial_R^{J-(l-1)} \left( R^{2m}\cosh^{1-l}R\, \sinh^{1+l}R \right) \right]_{R=0}.
\end{equation}
Notice that we have restricted $l$ to range from $0$ to $J+1.$ This is because the terms with $l>J+1$ all vanish at $x=R$. Now, (\ref{jth0}) can be written as 
\begin{equation}\label{feq}
\sum_{l}(l+1)!2^{l} \Delta^{(2l)}_{xx}(t_0,x_0)C^{J,l,0}(R=0) = -\sum_{l,m \neq 0}\frac{(l+1)!\,2^{l}}{(2m)!}\partial_x^{2m} \Delta^{(2l)}_{xx}(t_0,x_0)C^{J,l,m}(R=0).
\end{equation}
Setting $ J=0, $ we get 
\begin{equation}\label{J0}
\Delta_{xx}^{(0)}(t_0,x_0) = 0.
\end{equation}
$J=2$ gives 
\begin{align}\label{J2}
\nonumber \Delta_{xx}^{(2)}(t_0,x_0)   =& \   -\frac{1}{2}\Delta_{xx}^{(0)}(t_0,x_0) + \frac{3}{8}\partial_x^2\Delta_{xx}^{(0)}(t_0,x_0),\\
\Rightarrow \qquad \Delta_{xx}^{(2)}(t_0,x_0)   =& \   0.
\end{align}
$J=4$ gives 
\begin{align}\label{J4}
\nonumber \Delta_{xx}^{(4)}(t_0,x_0)   =& \   -\frac{2}{9}\Delta_{xx}^{(2)}(t_0,x_0) + \frac{1}{3}\partial_x^2\Delta_{xx}^{(2)}(t_0,x_0)-\frac{1}{9}\Delta_{xx}^{(0)}(t_0,x_0)\\
 & + \frac{5}{18}\partial_x^2\Delta_{xx}^{(0)}(t_0,x_0) + \frac{5}{144}\partial_x^4\Delta_{xx}^{(0)}(t_0,x_0)\\
\Rightarrow \qquad \Delta_{xx}^{(4)}(t_0,x_0)   =& \   0
\end{align}
Let's check it for $ J=2N $. It is easy to see that for all $ l \geq N+1 $ the coefficients $C$ are zero for all $m$. For $ l = N $ 
\begin{equation}\label{CN}
C^{2N,N,0}(R=0) = [\partial_R^{N+1}(\cosh^{1-N}R\sinh^{1+N}R)]_{R=0} = (N+1)!   \, .
\end{equation}
So the last non vanishing term in the $l$ series in (\ref{feq}) with $m=0$ is $ \Delta_{xx}^{(2N)}(t_0,x_0) $ and with $ m \neq 0 $ it will be a lower order term. Now notice that this term is a linear combination of all the lower order terms and their derivatives, which are zero. Hence $ \Delta_{xx}^{(2N)}(t_0,x_0) $ is also zero for $N=0,1,\ldots \ .$

 \subsection*{Numerical analysis for odd $ n $} 
 
\begin{figure}[h!]
\begin{minipage}{.5\textwidth}
\includegraphics[scale=0.5]{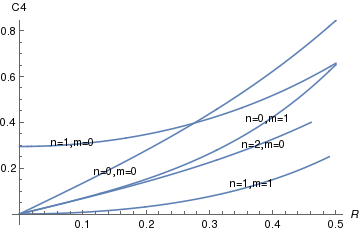}
\caption{$ C^4_{n,m} $ vs $R$ for different values  of $n$ and $m$.} 
\label{c4}
\end{minipage}
\begin{minipage}{0.5\textwidth}
\includegraphics[scale=0.5]{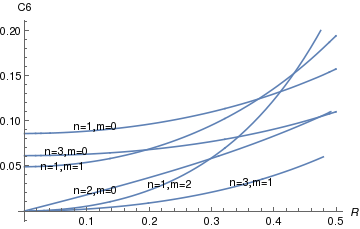}
\caption{$ C^6_{n,m} $ vs $R$ for different values of $n$ and $m$.}
\label{c6}
\end{minipage}
\end{figure}

Consider the expansion ($ \ref{exeqn} $): 
\begin{equation}
\sum_{n,m,j}\frac{2^{n+1}}{(2m)!}\partial_x^{2m} \Delta^{(n)}_{xx}(t,x_0)R^{j} C^j_{n,m}(0) = 0 ,
\end{equation}
where 
\[
C^j_{n,m}(0)=\frac{1}{j!}\left[\partial^j_R I_{n,m}(R) \right]_{R=0}.
\]
We have already seen that the first two terms in this series do not give any constraint and that the terms with odd powers of $ R $ (i.e, odd $j$) impose constraints on even $ n $. Thus one should expect that terms with alternative powers of $ R $ (i.e, even $j$) should impose constraints on odd $ n $. Let us check the term of order $ R^4 .$ Using numerics one can see that the coefficient $ C^4_{n,m}(0) $ is zero for all $ n,m $ except for $ n=1,m=0 $. The relevant plots are shown in figure $ \ref{c4} $. 

At order $ R^6 $, there are three non-zero coefficients, namely, $ C^6_{3,0},C^6_{1,0},C^6_{1,1} $. This implies that $ \Delta^{(3)}_{xx}(t_0,x_0) $ can be written in terms of $ \Delta^{(1)}_{xx}(t_0,x_0) $ and $ \partial_x^2\Delta^{(1)}_{xx}(t_0,x_0) $, and hence $ \Delta^{(3)}_{xx}(t_0,x_0) $ is also 0. Figure $ \ref{c6} $ shows the behavior of coefficients at $ R=0 $. 
 
\begin{figure}[h!]
\centering
\includegraphics[scale=0.63]{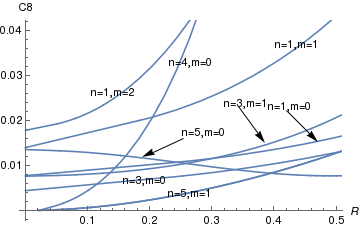}
\caption{$ C^8_{n,m} $ vs $R$ for different values of $n$ and $m$. }
These figures show for a particular $ j $ what coefficients $ C^j_{n,m} $ are non-zero as $ R\to 0 $.
\label{c8}
\end{figure}

 At order $ R^8 $, there are several non-zero coefficients which are shown in figure $ \ref{c8} $. From the plot, one can deduce that $ \Delta^{(5)}(t_0,x_0) $ can be written as a linear combination of the lower order terms and their derivatives, and hence  is 0. 
 
 From this pattern, we conclude that in general the higher order coefficients can be written in terms of lower order coefficients. This means that 
\[
\Delta^{(n)}_{xx}(t_0,x_0)=0 
\]
for all $ n $.

\newpage

\paragraph*{\bf Some numerical computations for boosted black brane:\\} 
 Here we have plotted different coefficients of $ R^4 $-th and $ R^5 $-th terms in the expansion series ($ \ref{bsum} $) around $ R=0 $. They are denoted by $ C^4_{\mu\nu} $ and $ C^5_{\mu\nu} $ respectively.

\begin{figure}[h!]
\begin{minipage}{.5\textwidth}
\includegraphics[scale=0.5]{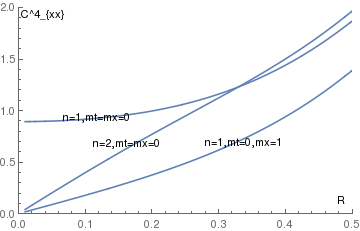}
\caption{$ C^4_{xx} $ vs $R$ for different values of $n, \ m_t $ and $m_x$.} 
\label{bc4xx}
\end{minipage}
\begin{minipage}{0.5\textwidth}
\includegraphics[scale=0.5]{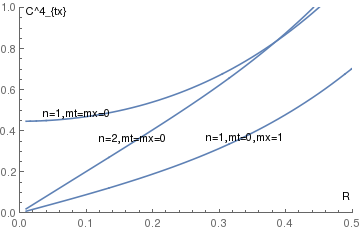}
\caption{$ C^4_{tx} $ vs $R$ for different values of $n, \ m_t $ and $m_x$.}
\label{bc4tx}
\end{minipage}
\end{figure}

\begin{figure}[h!]
\begin{minipage}{.5\textwidth}
\includegraphics[scale=0.5]{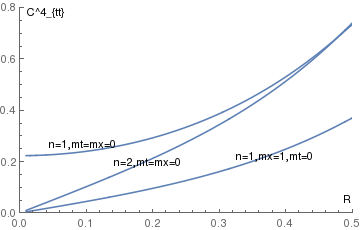}
\caption{$ C^4_{tt} $ vs $R$ for different values of $n, \ m_t $ and $m_x$.} 
\label{bc4tt}
\end{minipage}
\begin{minipage}{0.5\textwidth}
\includegraphics[scale=0.5]{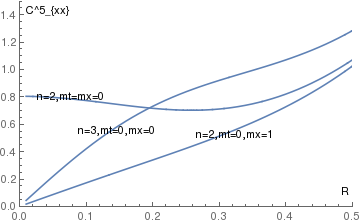}
\caption{$ C^5_{xx} $ vs $R$ for different values of $n, \ m_t $ and $m_x$.}
\label{bc5xx}
\end{minipage}
\end{figure}

\begin{figure}[h!]
\begin{minipage}{.5\textwidth}
\includegraphics[scale=0.5]{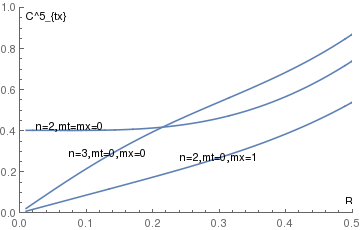}
\caption{$ C^5_{tx} $ vs $R$ for different values of $n, \ m_t $ and $m_x$.} 
\label{bc5tx}
\end{minipage}
\begin{minipage}{0.5\textwidth}
\includegraphics[scale=0.5]{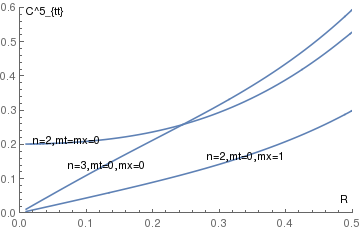}
\caption{$ C^5_{tt} $ vs $R$ for different values of $n, \ m_t $ and $m_x$.}
\label{bc5tt}
\end{minipage}
\end{figure}

Figures $ \ref{bc4xx} $, $ \ref{bc4tx} $ and $ \ref{bc4tt} $ show that all the coefficients of $ R^4 $-th term are zero as $ R\to 0 $ except when $ n=1 $ and $ m_t=m_x=0 $ which give the first equation in ($ \ref{beqns} $) with $ \beta=0.5 $. Similarly for $ R^5 $-th term the only non-zero coefficient with maximum value of $ n $  as $ R\to 0 $ is $ n=2 $ and $ m_t=m_x=0 $. This can be expressed as a linear combination of the terms with lower values of $ n $ which are already zero. These give the second equation in ($ \ref{beqns} $).


\end{document}